# Hype and Adoption of Generative Artificial Intelligence Applications

**Vinh Truong**
RMIT University
10 La Trobe Street, Melbourne, VIC, Australia
Email: vinh.truongnguyenxuan@rmit.edu.vn

## Abstract
New technologies often create opportunities for some while displacing others. On one hand, they enhance everyday life by supporting activities such as entertainment, education, and social connectivity. On the other, they increasingly replace humans in tasks that require productivity and analytical reasoning. The process of adapting to these technological shifts demands both technical adjustments and social readiness over time. It is widely recognized that for digital transformation initiatives to succeed, organizations and their workforce must first be psychologically prepared.

We are now entering the era of Generative Artificial Intelligence (AI), marked by the emergence of tools like ChatGPT, Bing AI, and Microsoft Office Copilot. Gaining insight into public sentiment and emotional responses toward these innovations is essential—not only for refining our theoretical models of technology acceptance but also for informing market strategies and adoption pathways. Drawing on the Gartner Hype Cycle and the Kübler-Ross Change Curve, this study posits that the adoption of generative AI applications follows a dual-stage process. It traverses the phases of the technology trigger, peak of inflated expectations, trough of disillusionment, slope of enlightenment, and plateau of productivity, while simultaneously reflecting the emotional stages of shock, denial, frustration, depression, experimentation, decision, and integration. To explore this hypothesis, the study employed sentiment and emotion analysis by collecting a large dataset of tweets related to generative AI. These were then quantified into sentiment and emotional scores to track user responses over time. While prior research has typically offered a binary snapshot of public sentiment at a single moment, this study takes a longitudinal approach—capturing the dynamic evolution of public attitudes. Unlike earlier studies, which presented raw data without theoretical linkage, this work bridges empirical evidence with established theoretical frameworks. Moreover, whereas past investigations largely focused on technology adoption from the perspective of information seekers, this study shifts the focus to content creators.

Given the recent public release of generative AI tools, there remains a significant gap in our understanding of how society is receiving and adapting to these innovations. As a result, policymakers face uncertainty about how to respond or how to guide markets in preparing for the changes these technologies will bring. Theoretically, this research contributes by empirically validating the applicability of the Gartner Hype Cycle and the Kübler-Ross Change Curve to the context of generative AI adoption. Practically, it offers valuable insights for businesses in planning the integration of generative AI tools and crafting supportive policies to enhance market readiness and resilience.

## 1. Introduction
Generative AI represents a form of artificial intelligence in which machines are trained on vast datasets—often labeled—and then use that learned knowledge to generate new content (Jovanovic, 2022). Unlike traditional AI, which typically classifies or predicts based on existing data, generative AI creates original outputs in response to user prompts. This includes tasks such as writing analytical reports, designing architectural plans, and generating programming code (Truong, 2024). Its capabilities far exceed those of earlier AI technologies, making it highly suitable for workplace integration and productivity enhancement—often performing tasks traditionally carried out by humans (Zhou et al., 2025).

Generative AI is no longer a future development—it is already available and actively used across various domains. Its growing influence is evident in the surge of user feedback across online platforms and forums. Unlike previous technologies primarily used for retrieving information, generative AI empowers users to *create* information (Freeda et al., 2024). This shift marks a fundamental transformation in how technology is integrated into both daily life and professional environments. The increasing adoption of generative AI in enterprises, especially when embedded into productivity suites like Microsoft 365 with Copilot, underscores its strategic importance and expanding reach (Liang et al., 2025).

However, this technological advancement has also sparked concerns—particularly among employees. As generative AI takes on roles traditionally performed by humans, many professionals worry about the security of their jobs (Sepehri et al., 2024). These concerns are no longer abstract; the effects are already being felt across diverse fields such as business analysis, software development, design, accounting, and investment. The question of whether generative AI will eventually replace human workers is becoming more urgent, especially as employees begin to interact directly with these tools. The introduction of generative AI thus evokes a mix of optimism and apprehension, especially within the context of employment (Sepehri et al., 2024). In addition to job-related anxiety, organizations also face challenges concerning data privacy, legal compliance, and ethical considerations—some even raising existential concerns about humanity's future alongside these powerful technologies (Dwivedi et al., 2023).

As generative AI applications became publicly accessible and entered workplaces, users began expressing their reactions—both emotional and cognitive—on social media platforms (Alhadlaq & Alnuaim, 2023). These responses offer a valuable window into how people are adjusting to the presence of generative AI. Understanding these public sentiments can help organizations reshape workplace strategies, restructure their workforces, and realign productivity goals to better navigate the evolving technological landscape.

While previous studies have extensively explored the adoption of new technologies, most have focused on constructs like perceived ease of use and perceived usefulness (Decaminada, 2022). Other research has examined technology adoption primarily from the perspective of information seekers, overlooking those who *create* information (Muriuki et al., 2024). This raises the question: should generative AI applications—tools for information creation—be assessed by the same criteria as earlier technological innovations?

To date, relatively little empirical research has examined generative AI adoption due to its recent introduction to public and business settings (Tuan et al., 2024). Additionally, existing sentiment analysis studies have largely provided snapshots of user sentiment at a single moment, failing to capture the evolving emotional journey over time. Some attempts have been made to track sentiment over time, but these often relied on raw data without anchoring in theoretical frameworks—limiting the generalizability of findings (Tuan et al., 2024). It is only recently that advances in machine learning have enabled more nuanced emotion classification, identifying up to 28 distinct emotions, which allows for a deeper understanding of how public opinions evolve (Demszky et al., 2020).

This study seeks to address these critical gaps by investigating the following research questions:
- **How have generative AI applications been hyped over time?**
- **How have generative AI applications been adopted over time?**

The remainder of this paper is organized as follows:
- **Section 2** presents a review of related literature and the theoretical models that inform this study.
- **Section 3** outlines the study's methodology, including data collection, analysis techniques, and tools used.
- **Section 4** summarizes the results of the data analysis and highlights key findings.
- **Section 5** discusses the broader implications of the findings, identifies limitations, and proposes directions for future research.

## 2. Literature Review

This study investigates public opinions and emotional responses to the hype and adoption of new technologies over time. It reviews key models such as technology life cycles and change curves, which serve as the theoretical foundation for analyzing related works. By comparing these frameworks with existing literature, the study identifies similarities, differences, and gaps in current research. These insights are then used to formulate the study's hypotheses.

### 2.1 Hype Cycle

In the current literature on technological innovation and adoption, recurring patterns emerge in how new technologies are introduced to the market and received by users. These patterns are often captured and visualized using conceptual models such as the S-curve and the Gartner Hype Cycle (Nikula, 2010; Shi, 2023). These models help illustrate the progression of technological development and user expectations over time, enabling researchers and practitioners to better understand the typical journey of new technologies from novelty to maturity.

The S-curve, originally proposed by Richard Foster and the McKinsey consulting group, describes a two-phase trajectory of technological development (Foster, 1986). It begins with a slow rate of improvement during the early stages of innovation, followed by a period of rapid progress as the technology matures and becomes more effective. Eventually, the pace of development slows again as the technology reaches saturation or technological limits. This model is largely linear and sequential, depicting a gradual shift from low to high maturity.

By contrast, the Gartner Hype Cycle offers a more nuanced view that incorporates not only technological maturity but also the psychological and emotional responses of users. This model emphasizes how excitement and disappointment evolve alongside the adoption process. The Gartner Hype Cycle combines two elements—a bell-shaped hype level and an S-shaped maturity level—into a five-stage model that explains how public perception shifts as technology moves from innovation to mainstream use (Kondo et al., 2022).

The five stages of the Gartner Hype Cycle are: (1) the technology trigger, (2) the peak of inflated expectations, (3) the trough of disillusionment, (4) the slope of enlightenment, and (5) the plateau of productivity. The cycle begins when a new technology first gains attention, often due to media coverage or promotional efforts. This early exposure is called the technology trigger, which sparks interest and curiosity among users and investors (Carr, 2017).

Following the initial trigger, excitement builds rapidly, leading to the peak of inflated expectations. At this stage, people often develop unrealistic hopes about the potential of the new technology, driven more by imagination than by actual use cases. Metrics such as social media engagement, attendance at conferences, and online searches surge during this time. The hype is further amplified by promotional content, leading users to believe in the technology's transformative potential (Oosterhoff, 2020).

However, once users begin to engage directly with the technology, a gap often emerges between their expectations and the actual performance or applicability of the product. This leads to the trough of disillusionment, a critical stage during which users express dissatisfaction or lose interest (Carr, 2017). The disappointment does not necessarily stem from the technology's failure but from overly optimistic expectations that were not grounded in reality.

Eventually, users begin to understand the realistic capabilities and limitations of the technology. This is the slope of enlightenment phase, during which people learn how to best use the technology in practical settings. Misconceptions are gradually replaced with informed opinions. As users adopt the technology in ways that align with its true strengths, its value becomes more evident and sustainable.

Finally, the plateau of productivity is reached. At this point, the technology gains widespread acceptance and is fully integrated into everyday practices. Its benefits are well-documented, and its use cases are understood. The hype has faded, but what remains is a stable, productive role for the technology in the market. Interestingly, in the Gartner model, the plateau of productivity is consistently higher than the starting point, reflecting genuine improvements over time (Dedehayir, 2016).

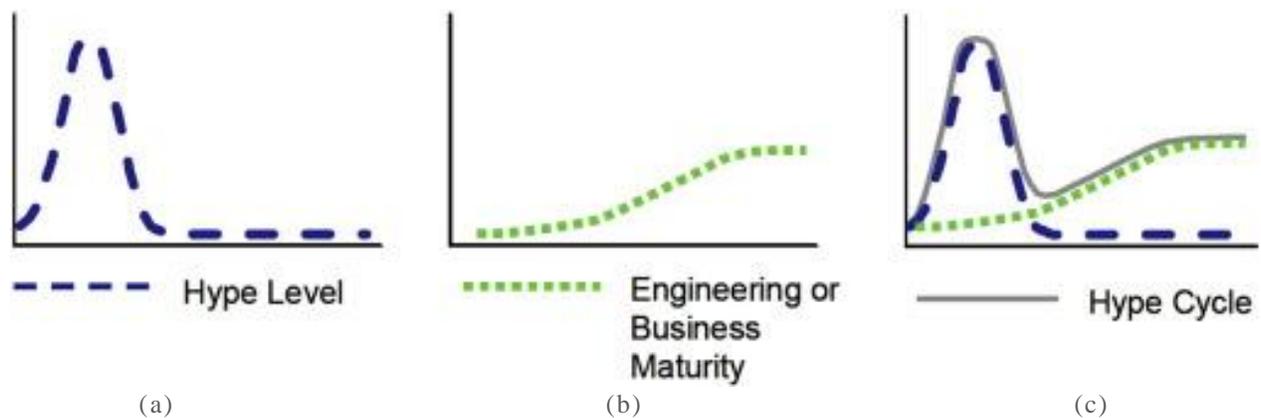

*Figure 1: Gartner Hype Cycle (Fenn & Raskino, 2008)*

The Gartner Hype Cycle has proven to be a useful tool for explaining and predicting public reception of various technologies across sectors such as education, healthcare, governance, and industry. For example, innovations like virtual reality, IoT platforms, and 4D printing have all been mapped onto this model to understand where they stand in terms of public interest and adoption (Truong, 2022). This helps organizations make better investment decisions based on realistic expectations.

One study, which analyzed 151 emerging technologies, demonstrated that most follow the general structure of the Gartner Hype Cycle, moving through all five stages before achieving widespread acceptance (Kaivo-Oja, 2020). However, the study also revealed that Generative AI was not included among the analyzed technologies, raising questions about whether this transformative technology will follow the same trajectory or deviate from it.

While earlier research suggests that user sentiment, especially as expressed on social media, fluctuates in line with the Gartner Hype Cycle, the duration and intensity of each stage can vary significantly by technology. Some innovations take years to move from the hype to the productivity stage, while others—particularly those adopted quickly due to urgency or trendiness—may complete the cycle within months or even weeks.

Generative AI presents a particularly interesting case. Unlike traditional hardware or software, it not only augments human capabilities but also raises concerns about job displacement, misinformation, and ethical dilemmas. The speed at which it has been adopted, coupled with its immense potential and risks, makes it unclear whether it will follow the conventional five-stage hype cycle or define a new path altogether.

Despite the frequent use of the Gartner Hype Cycle in academic and industry literature, few studies have systematically validated the model using real-time data. Some researchers have tried to infer hype stages by comparing technology rankings or measuring user engagement metrics, such as web traffic or event attendance (Decaminada, 2022). However, these proxies often fail to capture the underlying emotions and expectations of users at specific points in time.

More recently, machine learning and sentiment analysis tools have been used to better quantify user sentiment from social media data. By classifying posts as positive, neutral, or negative, researchers can begin to trace a more accurate emotional arc that aligns with the hype cycle (AlQahtani, 2021). However, a more refined sentiment analysis approach—such as continuous sentiment scoring—is needed to precisely map user emotions over time and correlate them with each stage of the hype cycle.

Taken together, the existing literature highlights both the value of the Gartner Hype Cycle in understanding the dynamics of new technology adoption and the gaps in empirical verification—especially for technologies as disruptive as Generative AI. This study, therefore, puts forward its first hypothesis:

**Hypothesis 1: The opinions about Generative AI applications will change through multiple stages following the Gartner Hype Cycle.**

## 2.2 Change Curve

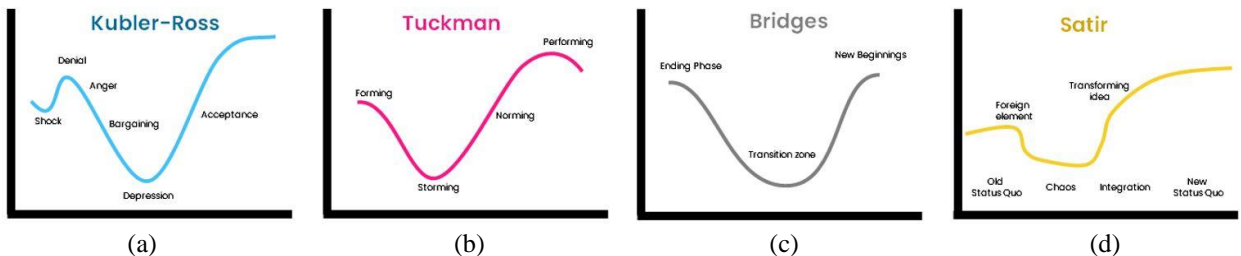

*Figure 2: Trend lines (Rogers, 2020)*

When new technologies are introduced in organizations or released into the broader market, they inevitably lead to transformations that extend beyond structural shifts and operational adjustments. While much attention in innovation adoption has traditionally focused on gathering user feedback or capturing general opinions, a more critical and often underexplored dimension lies in understanding emotions. Opinions are fleeting—shaped by transient thoughts or rational reflections—whereas emotions linger, influence decision-making over time, and determine the success or failure of technological integration. Emotions provide a deeper, more authentic representation of users' internal states during periods of change, making them essential to study in marketing, organizational behavior, and technology adoption (Truong & Hoang, 2022).

The process of technological adoption is fundamentally a process of change, and all change—whether welcomed or resisted—is inherently emotional. People associate changes in their environment with powerful emotional responses, including joy, excitement, anger, fear, and even grief. As the implementation of new tools or platforms often threatens the comfort of familiar routines, it triggers emotional upheaval. These emotional responses do not remain static; they evolve as individuals and teams progress through various stages of adaptation (Li et al., 2020). Understanding this emotional evolution allows organizations to better support employees and users throughout the transition.

Several well-established models in change management aim to capture the dynamics of transition. Tuckman's model of group development outlines the progression of team behavior through forming, storming, norming, and performing stages (Bonebright, 2010). Bridges' Transition Model focuses on the psychological reorientation individuals undergo, highlighting stages of letting go, the neutral zone, and a new beginning (Bridges, 2022). Similarly, the Satir Change Model portrays change as a nonlinear process with phases of resistance, chaos, integration, and stabilization (Workshop, 2018). Each of these models provides valuable insights into group and individual behavior, yet they often neglect the emotional underpinnings that shape these stages.

Among these models, the Kübler-Ross Change Curve stands out for its unique focus on emotions. Originally designed to explain the grieving process in terminally ill patients, it has since been widely adapted to understand emotional reactions to change in various contexts, including workplaces and technology adoption. Unlike models that emphasize structural stages or cognitive transitions, the Kübler-Ross model provides an emotional roadmap that tracks how individuals respond to disruptive news or events: starting from denial, progressing through anger, bargaining, and depression, and ultimately reaching acceptance (Williams, 2019).

When a change is introduced—such as a new AI system or software upgrade—users typically experience shock or denial as their initial reaction. This resistance stems from an attachment to the status quo, compounded by uncertainty about the new system's value and relevance. In this stage, individuals may freeze, ignore the implications of change, or outwardly deny its necessity. These reactions are not just cognitive disagreements but are deeply rooted in emotional discomfort and fear of the unknown.

As reality sets in and resistance becomes futile, users may enter the bargaining or frustration stage. Here, they attempt to negotiate their way back to familiar routines, questioning the need for change or suggesting partial implementations. This phase often gives way to a sense of helplessness and despair, especially when their efforts to retain the old system fail. The depression stage of the Kübler-Ross curve reflects the emotional low point, where individuals feel overwhelmed, defeated, and disconnected from the change process.

Despite its somber name, the depression stage is not a dead end. It represents a turning point. In the realm of technology and workplace change, most individuals do not remain stuck in this phase. With support, communication, and a clear demonstration of the new system's value, users begin to explore the new technology. They experiment, make sense of its features, and gradually develop the confidence to move forward. This marks the beginning of the acceptance and integration phase, where emotional resistance is replaced by cautious optimism.

This emotional journey applies not only to individual users but also to teams and organizations. Consider how employees respond to job reallocations or new digital tools. Resistance is often strongest among those with longer tenure or limited digital fluency. As observed in prior studies, older workers in particular associate new technologies with job insecurity and skill obsolescence. However, with time and appropriate training, these emotions evolve, mirroring the stages outlined by the Kübler-Ross model.

What makes the Kübler-Ross model especially powerful in technology contexts is its emotional vocabulary (Abebe, 2023). It does not merely describe abstract stages—it labels them with feelings like shock, denial, anger, sadness, and acceptance. This allows managers and researchers to frame resistance not as irrational behavior, but as a predictable, emotionally driven process. By validating these emotions and addressing them constructively, organizations can ease transitions and foster more empathetic change strategies.

However, there has long been a limitation in applying the Kübler-Ross model empirically: the difficulty in tracking and measuring emotions over time (Shi, 2023). Traditional methods of collecting feedback—surveys or interviews—often fail to capture the fluid and multidimensional nature of emotions. Emotional analysis was historically limited to six basic categories proposed by psychologist Paul Ekman: happiness, sadness, disgust, fear, surprise, and anger (Ekman & Friesen, 1969). These categories are simplistic and fail to capture the nuances of real emotional states in complex environments.

Advancements in affective computing and machine learning have begun to address this gap. Robert Plutchik's wheel of emotions expanded the vocabulary to eight primary emotions and their complex derivatives, providing a richer framework for emotional classification (Plutchik, 2001). More recently, researchers like Alan S. Cowen and Dacher Keltner have mapped out as many as 27 discrete emotional categories based on empirical data, offering a far more detailed emotional landscape (Cowen & Keltner, 2017). These insights make it possible to analyze emotional trajectories with unprecedented granularity.

Modern NLP techniques powered by transformer-based models such as BERT and its optimized variant RoBERTa have enabled the automatic classification of textual data—tweets, reviews, internal messages—into emotion categories (Liu, 2019). Tools like EmoRoBERTa leverage these advances to identify not just six, but dozens of emotional states with high precision. This breakthrough makes it possible to construct real-time emotional curves that reflect the Kübler-Ross model's trajectory, moving from denial and resistance to eventual acceptance (Kamath et al., 2022).

Despite these advances, the emotional impact of specific technological innovations—particularly those with job-disruptive potential like Generative AI—has yet to be deeply explored (McFarlane, 2024). Generative AI, with its capacity to automate content creation, legal writing, coding, and other high-skill tasks, represents a seismic shift in the employment landscape. Understanding how workers emotionally respond to this emerging technology is essential, not only for supporting workforce transitions but also for guiding ethical development and responsible implementation.

This gap in research forms the basis for the current hypothesis:

**Hypothesis 2: User and employee opinions about Generative AI will evolve through multiple emotional stages consistent with the Kübler-Ross Change Curve.**

## 3. Methodology

The research described is centered on understanding public sentiment and emotional responses to the introduction of generative AI applications, utilizing the Gartner Hype Cycle and the Kübler-Ross Change Curve as theoretical lenses. These two models are traditionally applied to track the adoption of technologies and human responses to change, respectively. By combining these frameworks, the study aims to investigate both the rational expectations and the emotional reactions that accompany the emergence of generative AI. Adopting a deductive research approach, the study begins with a set of hypotheses derived from these models and uses empirical data to either validate or challenge them. Importantly, the study's goal is not merely descriptive; it intends to interpret the collected data through these theoretical perspectives and contribute to the conceptual understanding of how society receives breakthrough technologies.

In terms of methodology, the study emphasizes the need to collect genuine, unsolicited user feedback, which traditional research methods such as in-person surveys or structured interviews may fail to capture. Social media platforms, particularly Twitter, provide an ideal environment for this purpose. As generative AI applications like ChatGPT, Bing AI, and Microsoft Office Copilot became publicly accessible, X (formerly Twitter) became a natural venue for users to voice their opinions and emotions. The open and unfiltered nature of this data source offers two main advantages: a broader reach that captures real-time feedback from a diverse user base, and the ability to access large volumes of public content without the constraints of sampling frames or recruitment.

To ensure robust and reliable data collection, the study leverages Twitter's public API, a widely recognized tool in academic research for accessing real-time data streams. A search query including terms like "ChatGPT," "Bing AI," and "Microsoft Office Copilot" was executed to retrieve relevant tweets. The collected data was then cleaned and preprocessed—duplicates, irrelevant hashtags, and empty fields were removed—to ensure quality. Each tweet includes a timestamp, which allows the researchers to track changes in sentiment and emotion over time. This temporal aspect is critical, as it supports the generation of dynamic charts that mirror the progressions outlined in the Gartner and Kübler-Ross models.

The first analytical stage involves sentiment analysis using a tool called VandeSentiment (Hutto & Gilbert, 2014). Unlike basic sentiment classifiers that label text as simply positive, negative, or neutral, VandeSentiment produces a compound score ranging from -1 to +1. This score reflects the intensity of sentiment, with values near +1 indicating strong positivity and those near -1 representing strong negativity. The advantage of using compound scores lies in the ability to produce time-series data, which can be plotted to generate sentiment curves. These curves can then be directly compared with the stages of the Gartner Hype Cycle—beginning with innovation triggers and moving through phases like peak of inflated expectations, trough of disillusionment, and plateau of productivity.

To explore emotional responses, the study incorporates EmoRoBerta, a sophisticated emotion detection model built upon RoBERTa and BERT, which are well-established natural language processing models (Kamath et al., 2022). Unlike traditional tools that identify only a handful of emotions, EmoRoBerta classifies texts into 28 nuanced categories. These categories range from primary emotions like joy and sadness to more complex ones such as admiration, embarrassment, and realization. This granularity enables the research to capture the full spectrum of emotional responses to generative AI technologies and map them onto the seven stages of the Kübler-Ross Change Curve, which include shock, denial, frustration, depression, experimentation, decision, and integration.

By applying EmoRoBerta to the same tweet dataset, the researchers calculate the total emotion scores per day for each emotion category. These scores are then plotted to form emotion curves, which visually represent the shifting emotional landscape of public reaction over time. This visual representation allows researchers to compare real-world emotional responses with theoretical expectations outlined in the Kübler-Ross Change Curve. For instance, a spike in emotions such as fear and confusion may coincide with the denial or frustration stages, while an increase in excitement and optimism may signal the transition toward experimentation or decision phases.

This dual-approach methodology—employing both sentiment and emotion analysis—strengthens the study's capacity to validate its hypotheses and contribute to theoretical development. Deviations from the expected curves, such as an extended period of frustration or a premature rise in optimism, will be critically analyzed. These deviations may reveal new insights into how users engage with rapidly evolving technologies, and they may necessitate revisions to the established models or the introduction of new conceptual constructs. Thus, rather than rigidly confirming existing theories, the study remains open to theoretical innovation, making room for dynamic and evolving interpretations of public engagement with AI.

## 4. Results

### 4.1 Sentiment Analysis

The graph in Figure 3 illustrates the sentiment scores of tweets mentioning generative AI over the first 100 days following the public release of ChatGPT. Rather than measuring interest solely by volume, this figure highlights the emotional tone and perception of the technology as reflected in user-generated content. While tweet volume helps identify spikes in attention, sentiment scores delve deeper into how people actually feel. The use of VADER sentiment analysis offers a nuanced understanding of public opinion by assigning a compound score to each tweet, ranging from -1 (most negative) to +1 (most positive). The overall average sentiment remains slightly positive, offering a meaningful contrast to fluctuations in tweet volume.

In the first 30 days, the sentiment score increased steadily and peaked around day 30 at approximately 0.37. This initial high reflects a wave of optimism and enthusiasm commonly seen during the early stages of the Gartner Hype Cycle—known as the "peak of inflated expectations." During this period, users were likely amazed by the new capabilities of ChatGPT and generative AI more broadly. Many tweets likely praised the technology's innovation, ease of use, or potential applications in education, writing, and coding. This peak coincides with the period when tweet volume was also high, reinforcing the sense of both excitement and curiosity.

Following the peak, the sentiment score experienced a noticeable decline, reaching its lowest point—around 0.24—roughly one month after the initial excitement. This drop suggests a shift toward disillusionment as users began to encounter limitations, errors, or ethical concerns associated with generative AI. This phase mirrors the "trough of disillusionment" in the Hype Cycle, where inflated expectations give way to more grounded critiques. While the number of tweets began to recover after day 30, sentiment scores suggest that the quality of opinions became more critical or skeptical, revealing growing concerns among users.

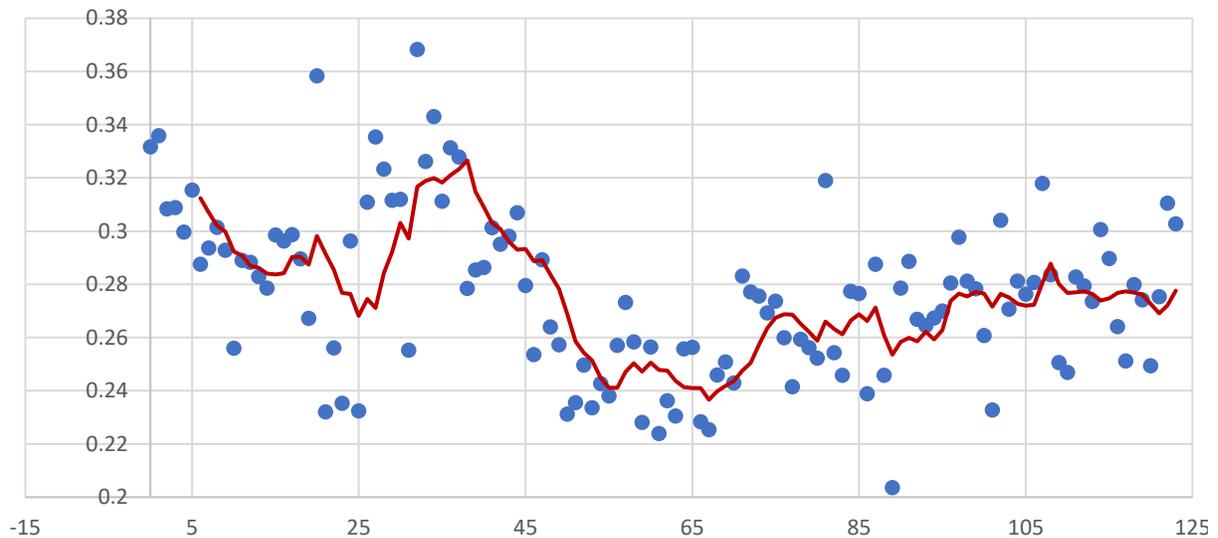

*Figure 3: Sentiment scores about GAI over time*

Despite this downturn, the recovery in sentiment was evident in the second and third months. The moving average line on the chart gradually ascends again, stabilizing around 0.27 by the third month. This signals a transition toward the "slope of enlightenment" and the "plateau of productivity," as users start to form more realistic assessments of the technology's value. At this point, adoption becomes more measured, and the emotional response more balanced. The fact that sentiment scores stabilize rather than continue rising suggests that while some enthusiasm persists, the initial hype has normalized into moderate optimism.

The movement of the sentiment scores closely follows the pattern predicted by the Gartner Hype Cycle. The bell-shaped curve and subsequent leveling off reflect a typical trajectory of new technologies—from rapid excitement to disappointment, followed by gradual recovery and stabilization. Importantly, the sentiment never turns strongly negative, indicating that while user perceptions dipped, they remained overall more positive than neutral. This subtle positivity is meaningful in technological diffusion, suggesting a successful but cautious integration of generative AI into users' expectations and daily lives.

What makes this sentiment-based analysis especially compelling is that it captures the psychological and emotional processing of users, something that raw tweet counts cannot do. When individuals post about a new technology, they are expressing not just interest, but hope, doubt, excitement, or even frustration. These emotional undertones are what shape public discourse and influence long-term adoption. In this context, the 0.27 average sentiment score by day 100 indicates a cautiously optimistic outlook—strong enough to support continued growth, but tempered by real-world concerns and limitations.

Interestingly, the sentiment score never fully recovers to its original peak after the downturn. This asymmetry suggests that while positive feelings return, the initial magic or novelty has faded. It aligns with the idea that first impressions of a groundbreaking technology can set unrealistically high expectations, which later have to be recalibrated. That recalibration is exactly what's shown in the sentiment curve, and its eventual stabilization represents a more sustainable understanding of the technology's role and value.

The relatively moderate sentiment score of 0.27 toward the end of the observation period implies that public opinion is neither polarized nor blindly optimistic. Instead, it reflects a nuanced perspective, one that acknowledges the promise of generative AI while also recognizing its flaws. The following emotional analysis—examining specific emotions such as admiration, confusion, or fear—will likely provide further clarity on what users appreciated or struggled with. These insights are essential for developers and policymakers seeking to improve AI services and manage user expectations effectively.

## 4.2 Emotional Analysis

The emotional analysis conducted through EmoRoBERTa reveals a multi-layered emotional journey that users experienced when interacting with generative AI applications. Table 1 provides insight into the evolution of twenty-eight emotions over several months, mapping them to the Kübler-Ross Change Curve. This analysis indicates that users' emotional reactions evolved from initial shock and denial to eventual acceptance and integration. The high emotional volatility in the early phase suggests the profound impact of generative AI on people's perceptions and expectations, not just cognitively but emotionally.

In the first phase of shock, surprise, joy, admiration, and pride were the dominant emotions. These peaked in the first month, indicating that users were amazed by the capabilities of generative AI. Surprise, in particular, had the steepest declining slope of -1.05, showing how rapidly this emotion faded as users became more familiar with the technology. Joy and pride, with positive slopes (0.38 and 0.34 respectively), suggest that while the initial astonishment decreased, users maintained a level of positive engagement. Admiration remained relatively stable (slope = 0.00), showing it neither grew nor diminished over time.

During the denial stage, emotions such as disgust, embarrassment, fear, and relief were most prominent. These emotions also peaked in the first month. The positive slope of disgust (0.47) is especially noteworthy as it indicates a growing repulsion, possibly driven by ethical concerns or unsettling responses from AI models. Relief (0.79), on the other hand, increased steadily, suggesting that despite initial concerns, some users found comfort or reassurance in the technology's consistency or usefulness. Embarrassment and fear also trended upward, pointing to unease or personal vulnerability in the face of such powerful systems.

The frustration stage followed in the second month, characterized by annoyance, anger, nervousness, and confusion. These emotions illustrate users' attempts to grapple with the limitations or unintended consequences of AI systems. For example, anger (slope = 0.04) and confusion (slope = -0.56) reflect users' struggle to understand how generative AI operates and how it fits into their lives or workflows. Nervousness and annoyance showed slight negative trends, indicating these feelings slightly subsided over time, perhaps as users gained more experience or understanding.

The sadness stage, which peaked around the third month, included emotions like sadness, disappointment, disapproval, and grief. These reflect a deeper emotional downturn, as users started to confront limitations, biases, or existential implications of generative AI. Sadness had a relatively steep negative slope of -0.63, highlighting a significant emotional dip. Disappointment and disapproval also had downward trends, signaling disillusionment. Interestingly, grief had a positive slope (0.14), suggesting a growing sense of loss or concern, potentially related to job security or human uniqueness being challenged by AI.

As users moved into the decision phase, they began to recognize the potential of generative AI beyond initial emotional turbulence. Emotions like gratitude, excitement, realization, and curiosity began to emerge and trend positively. Gratitude (0.57) and excitement (0.31) both indicate growing appreciation for AI's utility. Curiosity (0.24) shows that users became more inquisitive, eager to explore and understand AI applications. Realization (0.31) represents a turning point in perception, where users start to see AI not just as a novelty or threat but as a tool worth embracing.

*Table 1: Emotion trends*

|    | Emotion | Month of highest scores | Slope of trendline | Kübler-Ross stage |
|----|---------|-------------------------|--------------------|-------------------|
| 1  | surprise | 1 | -1.05 | Shock |
| 2  | joy | 1 | 0.38 | Shock |
| 3  | admiration | 1 | 0.00 | Shock |
| 4  | pride | 1 | 0.34 | Shock |
| 5  | disgust | 1 | 0.47 | Denial |
| 7  | embarrassment | 1 | 0.32 | Denial |
| 8  | fear | 1 | 0.41 | Denial |
| 9  | relief | 1 | 0.79 | Denial |
| 6  | annoyance | 2 | -0.07 | Frustration |
| 10 | anger | 2 | 0.04 | Frustration |
| 13 | nervousness | 2 | -0.08 | Frustration |
| 20 | confusion | 3 | -0.56 | Frustration |
| 17 | sadness | 3 | -0.63 | Sadness |
| 18 | disappointment | 3 | -0.05 | Sadness |

| 19 | disapproval | 3 | -0.28 | Sadness |
| 22 | grief | 3 | 0.14 | Sadness |
| 11 | gratitude | 3 | 0.57 | Decision |
| 12 | excitement | 3 | 0.31 | Decision |
| 21 | curiosity | 3 | 0.24 | Decision |
| 14 | realization | 3 | 0.31 | Decision |
| 15 | love | 4 | 0.20 | Experiment |
| 16 | optimism | 4 | -0.06 | Experiment |
| 23 | remorse | 4 | -0.05 | Experiment |
| 24 | desire | 4 | -0.07 | Experiment |
| 25 | amusement | 4 | 0.00 | Integration |
| 26 | neutral | 4 | 2.22 | Integration |
| 27 | approval | 5 | -0.95 | Integration |
| 28 | caring | 5 | -0.62 | Integration |

The experiment stage marks an even more rational and constructive engagement with AI. Emotions like love (0.20), remorse (-0.05), optimism (-0.06), and desire (-0.07) illustrate an emotional balancing act. Love and optimism show cautious hopefulness, while remorse and desire reflect thoughtful reconsideration of prior attitudes and a willingness to explore possibilities. These emotions, though modest in trend, point to a more stable and mature relationship with AI, one grounded in critical evaluation and personal adaptation.

The final stage—integration—saw the emergence of neutrality, amusement, approval, and caring. Neutrality had the steepest positive slope (2.22), indicating that users ultimately came to accept AI as a normalized part of their digital environment. Amusement remained flat (0.00), perhaps signaling that while some novelty remained, it did not grow. Approval (-0.95) and caring (-0.62), however, had negative slopes, which may indicate that while users accepted AI, they also began to disengage emotionally, treating it more as a utility than a transformative force.

Notably, the shift from highly charged emotions in the earlier months to more subdued or reflective emotions in later months supports the Kübler-Ross Change Curve's progression from resistance to acceptance. This progression mirrors real-life adaptation patterns, where individuals often react to change with initial resistance and then gradually move toward exploration and commitment. In the context of generative AI, this emotional trajectory underscores the psychological complexity of technological adaptation.

The clustering of emotions around each stage provides insight into how emotional responses can inform user experience strategies for AI developers and communicators. For instance, during the early phases, messaging might focus on building trust and addressing fears, while in later phases, it could highlight educational resources and opportunities for integration. Recognizing that users experience a wide range of emotions—often simultaneously—can help companies design more empathetic and supportive user journeys.

This emotional analysis adds a valuable dimension to traditional sentiment analysis by revealing the psychological depth of user interactions with generative AI (Truong et al., 2020). Emotions evolve in discernible patterns, reflecting broader psychological models like the Kübler-Ross Change Curve. This not only helps explain why people react the way they do but also offers actionable insights for managing public perception and adoption of new technologies. As users transition from shock and denial to decision and integration, their emotions shift from volatile to stable, from resistance to acceptance.

## 5. Discussion and conclusions

The findings presented in the previous section provide compelling evidence that both proposed hypotheses have been effectively validated. The observed trajectory of public opinion and emotional response toward generative AI technologies mirrors a predictable yet complex journey. While initial public opinion was marked by optimism and enthusiasm, this soon gave way to skepticism and decline. Conversely, emotional reactions began in a negative space—rooted in fear and uncertainty—before gradually becoming more accepting and positive. This contrast between cognitive opinion and affective emotion reveals a nuanced process of technological assimilation. When placed side by side, these models of opinion and emotion highlight the dual nature of public response to technological innovation, underscoring that early excitement often masks underlying shock and anxiety.

This pattern is consistent with the frameworks provided by the Gartner Hype Cycle and the Kübler-Ross Change Curve. These two models, though developed in entirely different domains, converge on a shared truth: that adaptation to change is rarely linear or entirely rational. The Gartner Hype Cycle illustrates how new technologies often enjoy a surge of inflated expectations before plunging into disillusionment and eventually finding a realistic plateau of productivity. The Kübler-Ross Change Curve, initially conceived to describe stages of grief, offers a psychological lens through which to view the emotional turmoil sparked by disruptive innovation. The moving charts in this study map closely onto both models, validating their continued relevance in an era marked by rapid and often unsettling technological change.

Indeed, the visual representations used in the study powerfully demonstrate that people often respond to new technologies with resistance before eventually adapting (Truong, 2023). As the charts show, the volume of negative feedback diminished over time, suggesting an increase in public comfort with generative AI applications. Despite genuine concerns about job displacement and ethical risks, users began to incorporate these tools into their routines. This progression—from fear to familiarity—illustrates a maturing relationship between society and technology. It also signals an opportune moment for businesses to consider full-scale implementation of generative AI in the workplace (Truong & Hoang, 2022). The predictive power of both the Gartner and Kübler-Ross models thus proves invaluable, not only in understanding public sentiment but also in guiding strategic deployment.

A particularly valuable theoretical contribution of this study lies in its application of the Kübler-Ross Change Curve beyond its original context. Traditionally used in clinical psychology to describe the emotional journey of grief, the model has since been adapted to explain organizational change. However, it has rarely been applied to assess the psychological impact of technological disruptions. This study fills that gap, demonstrating that new technological developments can trigger a cascade of emotional reactions—ranging from shock and denial to frustration, despair, and ultimately, acceptance. By using a psychological model to explain reactions to innovation, the study highlights how the emotional toll of disruption must be factored into discussions about technology adoption and workforce readiness.

The successful adaptation of these models to a technological context not only enhances their theoretical robustness but also offers practical insights. Technological advancement is not solely a matter of efficiency or productivity; it is also a deeply human process that affects emotions, behaviors, and social dynamics. The fact that acceptance does not occur uniformly or immediately across a population is particularly important. As the study indicates, the general market cannot be considered fully receptive to a new technology until a critical mass of individuals has emotionally and cognitively accepted its presence. Organizations and policymakers must therefore develop strategies that facilitate not just technical integration but also emotional adaptation.

Despite these important contributions, the study is not without limitations. One key constraint is its general focus on generative AI without accounting for contextual differences across industries or technological ecosystems. For instance, how might public opinion and emotional response differ in sectors where generative AI is deeply integrated, such as healthcare, creative industries, or the Internet of Things? (Truong, 2016) Future research could apply the Gartner Hype Cycle and Kübler-Ross Change Curve in more targeted scenarios to test their explanatory power under specific circumstances. Moreover, as generative AI technologies continue to evolve, new emotional and behavioral patterns may emerge, requiring further theoretical refinement and empirical testing.

In summary, this study provides both a conceptual and practical framework for understanding how people respond to disruptive technologies like generative AI. By aligning emotional and cognitive responses with well-established models of change and adaptation, it offers a holistic perspective on technology adoption. For practitioners, especially marketers and organizational leaders, these insights are invaluable. They provide a roadmap for managing user expectations, guiding communication strategies, and designing adoption plans that acknowledge both the rational and emotional dimensions of change. Ultimately, the study underscores the importance of balancing innovation with empathy, ensuring that technological progress does not come at the expense of human well-being.